\documentclass[twocolumn]{svjour3}          
\smartqed  
\usepackage{graphicx}
\usepackage{array}
\usepackage{rotating}
\usepackage{multirow}
\usepackage{amsmath}

\usepackage{kantlipsum,widetext}

\AtBeginDocument{\setlength\abovedisplayskip{4pt}}
\AtBeginDocument{\setlength\belowdisplayskip{4pt}}

\usepackage{color}
\definecolor{naranja}{rgb}{1, 0.6, 0.2}
 
\begin{document}

\title{Immersion Metrics for Virtual Reality}

\author{Matias N. Selzer*\thanks{* Corresponding author.} \and Silvia M. Castro
}


\institute{Matias N. Selzer, and Silvia M. Castro \at
              Institute for Computer Science and Engineering (UNS–CONICET), and VyGLab Research Laboratory (UNS-CICPBA), Department of Computer Science and Engineering, Universidad Nacional del Sur, Bah\'{i}a Blanca, Argentina.\\
              \email{matias.selzer@cs.uns.edu.ar, smc@cs.uns.edu.ar}\\           
}

\date{Received: date / Accepted: date}

\maketitle

\begin{abstract}
Technological advances in recent years have promoted the development of virtual reality systems that have a wide variety of hardware and software characteristics, providing varying degrees of immersion.
Immersion is an objective property of the virtual reality system that depends on both its hardware and software characteristics.
Virtual reality systems are currently attempting to improve immersion as much as possible. 
However, there is no metric to measure the level of immersion of a virtual reality system based on its characteristics.
To date, the influence of these hardware and software variables on immersion has only been considered individually or in small groups. 
The way these system variables simultaneously affect immersion has not been analyzed either.
In this paper, we propose immersion metrics for virtual reality systems based on their hardware and software variables, as well as the development process that led to their formulation.
From the conducted experiment and the obtained data, we followed a methodology to find immersion models based on the variables of the system.
The immersion metrics presented in this work offer a useful tool in the area of virtual reality and immersive technologies, not only to measure the immersion of any virtual reality system but also to analyze the relationship and importance of the variables of these systems.

\keywords{Virtual Reality \and Immersion \and Presence}
\end{abstract}

\section{Introduction}
\label{sec:intro}
Virtual Reality (VR) systems are sophisticated human~-computer interaction interfaces that are used today in a wide variety of application areas such as education \cite{freina2015literature,monahan2008virtual,messinis2010investigation,qiao2021integration}, medicine \cite{ang2021recent,swiatek2021covid} and training \cite{grantcharov2004randomized,barbosa2017multisensory,aim2016effectiveness,joshi2021implementing}, among others.
Some of the most popular VR application areas today include entertainment and video games~\cite{zyda2005visual,cheok2009mixed,jasmine2021augmented} and each application has different objectives, requiring different hardware and software implementations.

For decades, it has been discussed in the literature which variables of a VR system are related to the perceived level of presence and immersion.
The influence of these hardware and software variables on immersion has only been considered individually or taking into account small groups of these.
To date, the way they all simultaneously affect immersion has not been analyzed. 
Also, the influence of all these variables have not been compared with each other.
This motivates the generation of metrics developed to calculate the perceived immersion of a VR system that include all its variables with their respective levels of incidence.

The main contribution of this work is the design and development of immersion metrics that calculate the level of immersion of a given VR system, based on its hardware and software characteristics.
This can be also used to compare the immersion of different commercial or \textit{ad hoc} VR systems. 
In addition, these metrics can be considered an extremely useful design tool to measure the immersion of prototypes, allowing to adjust, as much as possible, the values of the variables involved in the system. 
To achieve this goal, we followed the methodology presented in section \ref{sec:methodology}.

\section{Background and Related Work}\label{sec:background}

\subsection{Presence and Immersion}
\textit{Immersion} is a relevant concept in VR that has generated a lot of confusion regarding its similarity to the concept of \textit{presence}.
The feeling of presence is a subjective measure that depends on the sensation and personal experience of each user.
On the contrary, according to Slater et al. \cite{slater1999measuring}, immersion refers to an objective characteristic of a virtual environment that is strongly linked to both hardware and software components.
According to this, the wider the sensory bandwidth of a system, the more immersive the system would be.
For example, a system that includes 3D spatial sound should be more immersive than a system that does not include sound at all, or a system with a field-of-view (FOV) of 150º should be more immersive than a system with a FOV of 100º.
Also, in this context, two different users experiencing exactly the same VR system should perceive the same level of immersion.
However, this is not as simple as just increasing the variable values as much as possible. 
If we consider the sound volume, for example, a higher value (i.e. a louder volume) might not always produce a higher level of immersion.

In this work, we study the relationship between variables and follow Slater's definition of immersion, i.e., immersion is considered as an objective characteristic of a VR system that can be measured.

\subsection{Measuring Immersion}
There are many questionnaires and surveys to measure presence and immersion through causal factors and different variables. 
However, only a small number have been validated and are used regularly.
In 2004, Baren and Ijsselsteijn \cite{van2004measuring} presented a complete list of existing measurement methods, although today this list is out of date.

One of the most used tools to measure presence and immersion in virtual environments is the \textit{questionnaire}.
Each type of questionnaire has its advantages and disadvantages.
While questionnaires with many items can provide a detailed assessment of multiple dimensions of presence, single-item questionnaires, such as the test presented by Bouchard et al.~\cite{Bouchard2004-hw}, allow a rapid assessment and are less prone to memory impairment after exposure to the experience.
The Bouchard test has been used successfully in previous works \cite{gromer2018height,oberdorfer2018effects,selzer2019effects}.
In this work, since the user must perform as many trials as possible (see section \ref{sec:experiment}), we required a questionnaire that was easy to understand and quick to complete.
For this reason, in a similar manner as the Bouchard test, we used a single-item questionnaire for immersion.

\subsection{Variables Contributing to Immersion}
The literature presents an extensive work related to the variables that may contribute to a higher sense of presence in VR.
This relates to the characteristics of the user and those of the system.
The user characteristics refer to the psychological and subjective characteristics that influence the degree of perceived presence, and those of the system refer to the technical characteristics of the system that influence the perceived level of immersion.

Previous works present several variables related to the immersion and the visual features provided by the system.
These include the field-of-view \cite{kim2014effects,wallis2013predicting}, the screen resolution \cite{kim2014effects,ahn2014effects}, the stereopsis \cite{kim2014effects,ahn2014effects}, the response time or latency \cite{kim2020multisensory}, brightness, contrast, saturation, and sharpness \cite{Tomasz_Mazuryk1996-sy}, the level of detail of the 3D models \cite{volkmann2020you}, the lighting of the virtual environment \cite{slater2010simulating}, and the use of dynamic shadows \cite{slater2010simulating}.
Regarding the variables related to audio, these include the use of sound compared to not using sound \cite{zeltzer1992autonomy,poeschl2013integration}, the ambient sound \cite{Balakrishnan2011-yc}, the 3D spatial sound \cite{azevedo2014combining,bergstrom2017plausibility}, the use of headphones compared to the use of speakers \cite{bergstrom2017plausibility}, and the echo or reverberation \cite{Balakrishnan2011-yc}.
Finally, regarding the variables related to the user's tactile system and tracking, these include the sensory bandwidth \cite{snow1998charting}, the level of body tracking~\cite{gorini2011role}, the degrees of freedom \cite{Balakrishnan2011-yc}, the affordance of the controls \cite{williams2013effects}, the response time or latency of the tracking \cite{arthur1993evaluating}, the locomotion mode used to navigate through the virtual environment \cite{selzer2018interaccion}, and the temperature and wind \cite{azevedo2014combining}.
The supplementary material presents the complete and extensive list of variables studied for this work.

\begin{figure}  
\centering
\includegraphics[width=1\linewidth]{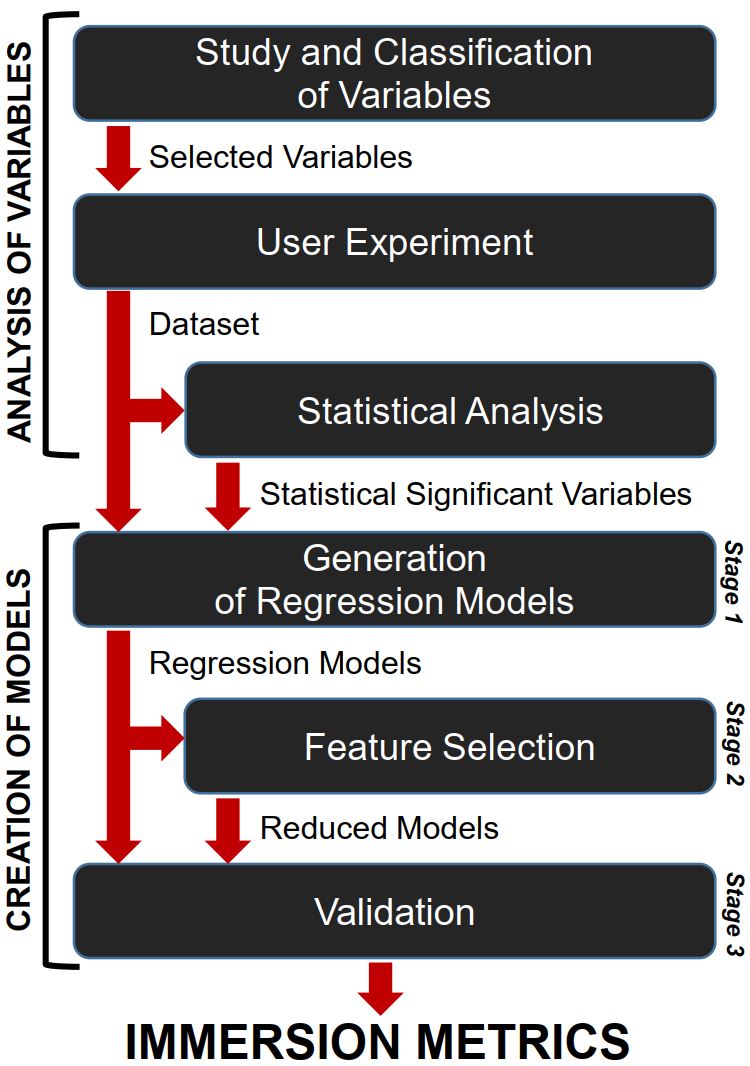}
\vspace{5pt}
\caption{
Methodology followed in this work to generate immersion metrics based on the hardware and software characteristics of the VR system. The first part deals with the analysis of the variables and the population of a dataset in a user experiment. The second part deals with the creation of immersion models by using techniques of linear regression, feature selection and validation.
1}
\label{fig:methodology}
\end{figure}

\section{Methodology}\label{sec:methodology}
In order to generate an immersion metric, we followed a methodology that can be divided into two main parts (see Fig. \ref{fig:methodology}).
The first part deals with the analysis of variables and the dataset population. 
It is very common for some variables to be named differently in different studies. 
Hence, the first step in this part of the methodology is the study and classification of all these variables.
The variables selected for the experiment are presented in section \ref{sec:var_independent}.

Once the variables are selected, we required a method to quantify the level of immersion produced by a VR system, for the different values that the variables can take.
For this reason, we designed a user study in which the user explores and interacts with a virtual environment and reports the level of perceived immersion. 
In each trial, this virtual environment is generated based on the values taken by the variables of the system.  
Hence, each trial contributes to a new sample in a dataset that stores the relationship between the VR system variables and the immersion perceived by the user. 
After that, statistical analyses were performed to this dataset to find which variables are statistically significant.
This process is detailed in section \ref{sec:datasetPopulationAndAnalysisofVariables}.

The second part of this methodology deals with the generation of immersion models, and is divided into 3 stages.
In Stage 1, different regression models for immersion are generated, based on the dataset obtained in Part 1 and the statistically significant variables.
This is detailed in section \ref{sec:stage1}.
Some models, for example, considered all the variables of the experiment and others considered only the statistically significant variables.
In Stage 2, feature selection techniques were applied to reduce the number of variables of the models.
This process is explained in section \ref{sec:stage2}.
All the candidate models (the models generated in Stage 1 and Stage 2) went through a validation process in Stage 3 (see section \ref{sec:stage3}).

Finally, the best models in terms of predictive power, number of terms and coefficients were selected as immersion metrics.
This is explained in detail in section \ref{sec:modelosSeleccionados}.

\section{Dataset Population and Analysis of Variables}\label{sec:datasetPopulationAndAnalysisofVariables}

\subsection{User Study}\label{sec:experiment}
We designed an experiment in which the participants have to perform a certain task within a virtual environment and report the perceived level of immersion.
That virtual environment is detailed in section \ref{sec:virtualEnvironment} but its visual, auditory and tactile composition depends on the values taken by the independent variables in each trial.
That is, each time a new trial is run, the independent variables take a new value, modifying the virtual environment completely (see section \ref{sec:procedure}).

\subsubsection{Participants}\label{sec:participants}
Initially, this experiment was intended to be performed by as many participants as possible, each one performing as many trials as possible, in order to populate a dataset. 
In that scenario, with a reasonably big dataset, more accurate analyses could be performed related to the variables and the differences between the participants' characteristics.
However, due to the conditions of social isolation related to COVID-19, we decided to carry out the experiment focused on a single user.
This type of experiment is better known as \textit{single-subject design} and is often used in fields such as psychology, education, and human behavior.
It is also often used to assess the effect of a variety of applied research interventions.

The present experiment was conducted by a 30-year-old male self-perceived gender participant.
The participant had experience playing video games and using VR systems.
Although the developed methodology has been used in the context of \textit{single-subject design}, it constitutes the basis for use with multiple users (see section \ref{sec:discussion}).

\subsubsection{Hardware}\label{sec:hardware}
The experiment was conducted using a desktop computer with an i5-7500 3.40GHz CPU, with 16GB of RAM, and a GeForce GTX 1060 6GB GPU video card.
There was no performance degradation that could have compromised the experience.
Visual stimulation and interactions were carried out using the Oculus Rift CV1 \footnote{https://www.oculus.com/rift} system.
The binocular field-of-view of the system is approximately 110º.
Its display has a 60Hz refresh rate and a resolution of $2160 \times 1200$ for both eyes.
Head orientation and position are recorded by the system's integrated gyroscope and accelerometer.
The optical cameras of the system were used to track the participant.
The system also has a mechanism to adjust the participant's visual disparity.
Finally, the system's integrated headphones were used to deliver the audio.

\subsubsection{Virtual Environment}\label{sec:virtualEnvironment}
According to Makransky et al. \cite{makransky2019adding}, to obtain the most accurate measurement possible, the participant must remain entertained and motivated throughout the experience.
If the participant becomes bored and begins to ramble, this can negatively influence the accuracy of the results.
For this reason, we designed a game to keep the participant motivated during the test.
In this game, the participant must survive a zombie attack for a certain period of time.
To keep the user motivated, the difficulty of the scenario varies depending on the remaining playing time.
That is, the frequency with which new enemies appear and their speed increase as time goes by.

\begin{figure*}  
\centering
\includegraphics[width=0.49\linewidth]{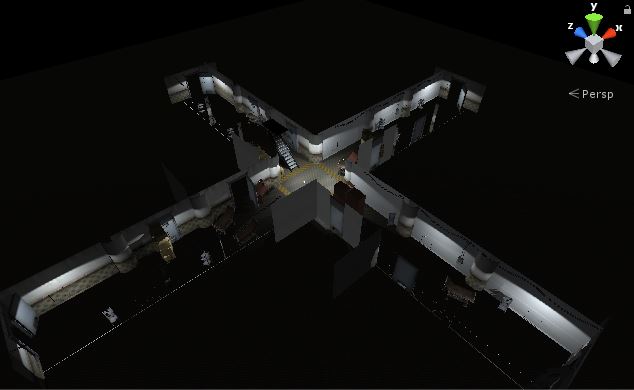}
\includegraphics[width=0.49\linewidth]{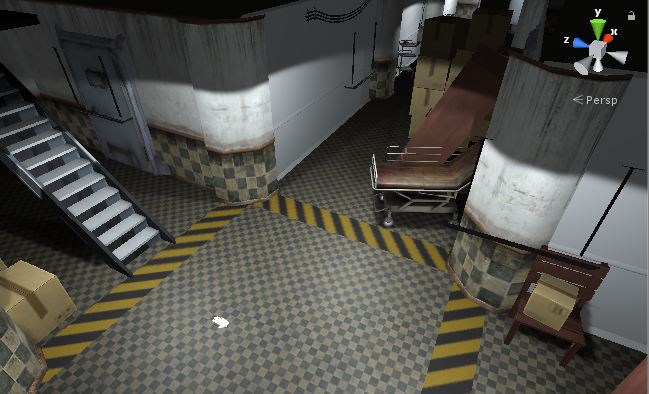}
\caption{
Top aerial view of the virtual environment (left) and a close-up view (right). The user is at the center. Enemies emerge from the 4 corridors' ends and walk towards the center.}
\label{fig:topView}
\end{figure*}

The participant is located at the center of a crossroad between two corridors.
The enemies appear at the end of those corridors and start walking towards the participant, who can only walk through a delimited (virtual) zone of $ 3m \times 3m $ (figure \ref{fig:topView}).
If the enemies get too close to the participant, the game ends.

To evaluate the participant's movements, different obstacles were placed to obstruct the vision between the participant and the enemies.
Hence, the participant needs to move to shoot the enemies.

\begin{figure*}  
\centering
\includegraphics[width=0.49\linewidth]{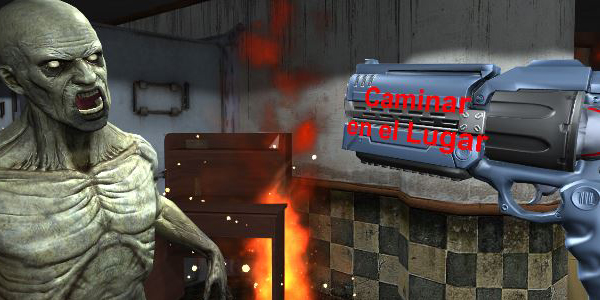}
\includegraphics[width=0.49\linewidth]{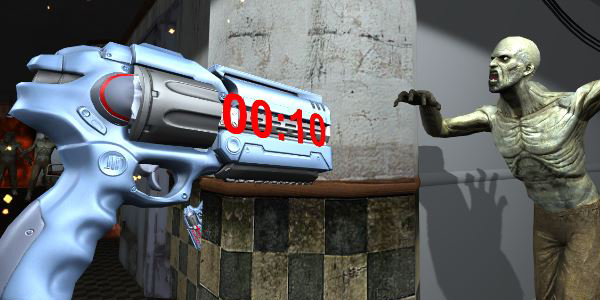}
\caption{
The left side of each gun shows the the walking mode (left).
The right side of each gun shows the remaining playing time (right).
}
\label{fig:armas}
\end{figure*}

The participant has a gun in each hand to shoot the enemies (figure \ref{fig:armas}).
The bullets are unlimited.
The right side of the guns shows the remaining time and the left side the locomotion mode.

The delivered audio includes other sounds in addition to the ambient background sound.
When the participant shoots, a shooting sound is generated from the gun.
In addition, the enemies produce three different sounds: a sound when they appear at the end of a corridor, another sound when they are close to the participant, and another sound when they die.

\subsubsection{Independent and Dependent Variables}\label{sec:var_independent}
The independent variables are those established by the system in each test and do not depend on other variables.
The variables considered for the experiment are listed in table \ref{table:variables}.

\begin{table*}
\centering
\caption{Independent variables considered in this study. These variables are arranged in categories, namely: Trial Variables, Visual Configuration Variables, Audio Configuration Variables and Locomotion Configuration Variables. For each variable, a brief description is presented.}
\resizebox{1\linewidth}{!}{
\begin{tabular}{cll}
\hline
\textbf{Category}                                                                                  & \multicolumn{1}{c}{\textbf{Variable Name}} & \multicolumn{1}{c}{\textbf{Description}}                                                     \\ 
\hline
\begin{tabular}[c]{@{}c@{}}\textbf{Trial}\\\textbf{Variables}\end{tabular}                         & Duration Time                              & from 120 to 1200 seconds (2 to 20 minutes)                                                   \\ 
\hline
\multirow{16}{*}{\begin{tabular}[c]{@{}c@{}}\textbf{Visual }\\\textbf{Configuration}\end{tabular}} & Screen Resolution (Width and Height)       & from 0.1 to 1.0 multiplied by the device max resolution (2160x1200 for the Oculus Rift CV1)  \\
                                                                                                   & Field-of-View (FOV)                        & from 30\% to 100\% of the device max FOV                                                     \\
                                                                                                   & Frame Rate (FPS)                           & from 8 to 60 FPS                                                                             \\
                                                                                                   & Stereopsis                                 & Enabled or Disabled                                                                          \\
                                                                                                   & Antialiasing (MSAA)                        & Enabled or Disabled                                                                          \\
                                                                                                   & Textures                                   & Enabled or Disabled                                                                          \\
                                                                                                   & Illumination                               & Ambient Light with No Shading, or Point Lights with Realistic Shading                        \\
                                                                                                   & Saturation                                 & from -1.0 (no saturation at all) to 1.0 (extremely saturated image)                          \\
                                                                                                   & Brightness                                 & from -0.8 to 0.8. Higher or lower values create completely dark or white scenes              \\
                                                                                                   & Contrast                                   & from -0.8 to 0.8                                                                             \\
                                                                                                   & Sharpness                                  & from 0.0 to 1.0                                                                              \\
                                                                                                   & Shadows                                    & Shadow Strength from 0.0 to 1.0                                                              \\
                                                                                                   & Reflections                                & (Specular Coefficient of Materials) Enabled or Disabled                                      \\
                                                                                                   & 3D Models Detail                           & Low-Poly Models or High-Poly Models                                                          \\
                                                                                                   & Depth-of-Field                             & Enabled or Disabled                                                                          \\
                                                                                                   & Particles                                  & Enabled or Disabled                                                                          \\ 
\hline
\multirow{4}{*}{\begin{tabular}[c]{@{}c@{}}\textbf{Audio }\\\textbf{Configuration}\end{tabular}}   & Sound System                               & No Sound, Speakers, or Headphones                                                            \\
                                                                                                   & Ambient Sound                              & Enabled or Disabled                                                                          \\
                                                                                                   & Reverberation                              & Enabled or Disabled                                                                          \\
                                                                                                   & 3D Spatial Sound                           & Enabled or Disabled                                                                          \\ 
\hline
\begin{tabular}[c]{@{}c@{}}\textbf{Locomotion}\\\textbf{Configuration}\end{tabular}                & Locomotion Mode                            & Real Walking, Teleportation, Joystick Movement, or Walking-in-Place (WIP)    \\
\hline
\end{tabular}
}\label{table:variables}
\end{table*}

On the other hand, the dependent variables are those that depend on the independent variables.
These variables are rated, on a scale from 1 to 100, with a questionnaire at the end of each test.
As in the Bouchard questionnaire \cite{bouchard2008anxiety}, we measured immersion with a specific question: ``How much did you feel immersed in the experience? i.e., how much did you feel that you SAW, HEARD and NAVIGATED like you do in real life?''.
The participant was given a thorough explanation on the question and also the opportunity to ask questions.

\begin{figure*}  
\centering
\includegraphics[width=0.47\linewidth]{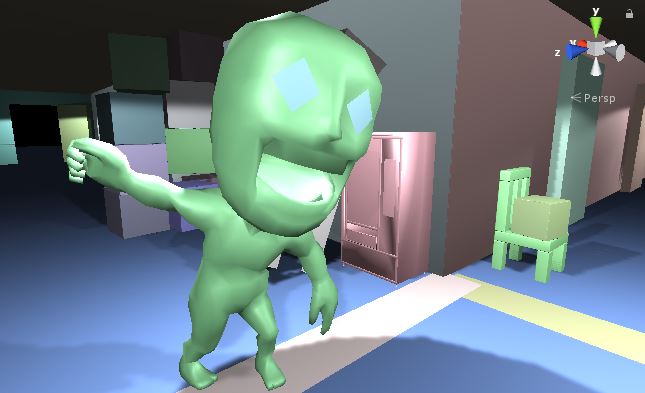}
\includegraphics[width=0.515\linewidth]{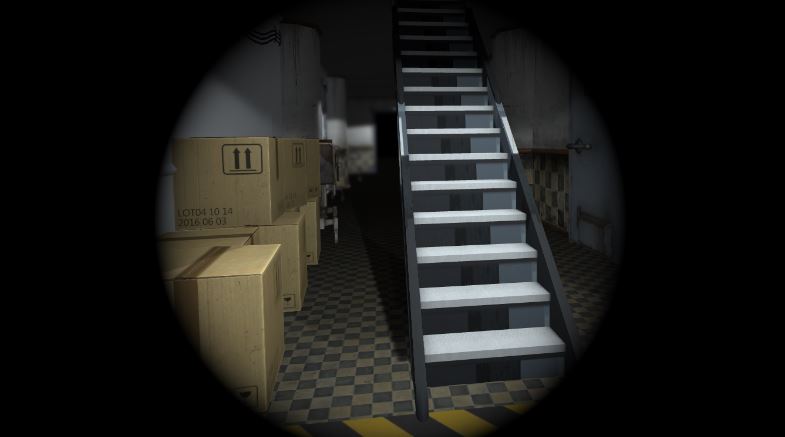}
\caption{
Scene with a low value of 3D models detail, no textures and a large field-of-view (left). Scene with a high value of 3D models detail, textures activated, and a narrow field-of-view (right).
}
\label{fig:textures}
\end{figure*}

\subsubsection{Procedure}\label{sec:procedure}

Each time a new trial begins, the characteristics of the scene related to all the independent variables are modified.
For numerical variables, a random real value is computed within the established range and, for categorical variables, a random integer value associated with one of the categories of that variable is computed.
The virtual scenario is then generated based on these variables and their computed values.
Hence, for each trial, the participant would perceive a completely different experience.
Figure \ref{fig:textures} shows two examples of different dynamically generated virtual scenes.
The supplementary material presents examples of the effect of the other variables on the virtual scene.

Each trial ends either when the participant survives for the specified time or when an enemy gets close enough.
Following the principles proposed by Slater et al. \cite{slater2002presence}, it is important to take measurements as soon as possible after the experience.
Immediately after the trial ends, the enemies that are still in the scene disappear and a floating screen appears for the user to answer the question related to the perceived total immersion.

Finally, it is important to mention that the participant took a 5-minute break between trials.
No noticeable symptoms of cybersickness occurred at any time.

\subsubsection{Results}\label{sec:exp-results}
The data from the experiment was saved into a dataset for later analysis.
This dataset is represented by a table, where each row corresponds to a sample and each column to a variable.
The data collected during each trial constitutes a sample in this dataset.
For this experiment, the participant performed 401 successful trials, thus generating 401 rows in the dataset.
The dataset is public and available online \cite{selzerDataset}.

\subsection{Statistical Analysis}\label{sec:dataAnalysis}
Based on the obtained dataset, we performed statistical analyses to evaluate the relationship between the different variables and the perceived immersion.
We present the most relevant results of the analyses relating total immersion.

We performed Kolmogorov-Smirnov tests for normality, which showed that the data did not follow a normal distribution. 
For this reason, we used non-parametric tests for statistical analysis, i.e., we employed non-parametric Kruskal-Wallis tests to evaluate the statistical differences of the independent variables on immersion.
We used Dunn’s pairwise comparison with Bonferroni correction to identify where the differences occurred.
In all these cases, a confidence interval of 95\% was considered.
Finally, correlation analyses were performed to study possible relationships between the independent variables and the perceived immersion.
We used Spearman correlations for ordinal variables and Pearson correlations for continuous variables.

\begin{figure*}  
\centering
\includegraphics[width=0.49\textwidth]{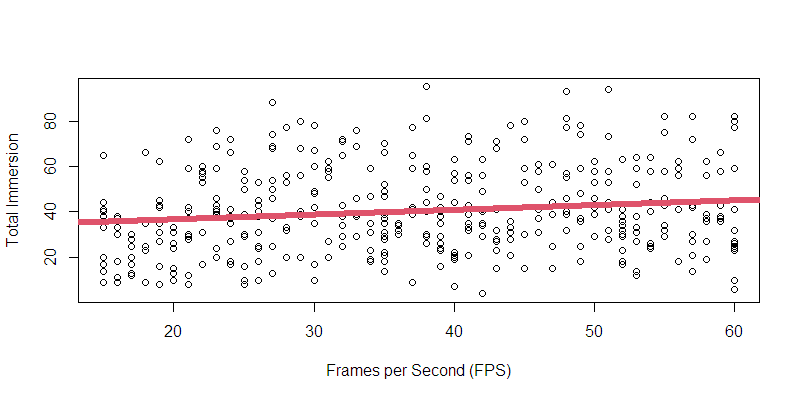}
\includegraphics[width=0.49\textwidth]{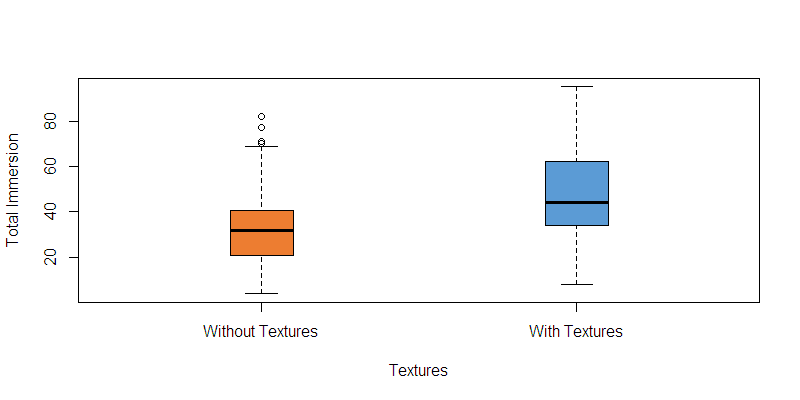}
\caption{
Relationship between frames per second and total immersion ($ r (401) = 0.148 $, $ p <0.01 $) (left). Boxplot for the relationship between the use of textures and total immersion (\textit{$ \chi ^ 2 = 65.017 $, $ p <0.01 $}) (right).
}
\label{fig:plotTotalvsTexturas}
\end{figure*}

Considering the visual variables with respect to total immersion, a small correlation was found with screen width ($ r (401) =  0.276$, $ p < 0.01$), frames per second ($ r (401) = 0.148 $, $ p <0.01 $) (figure \ref{fig:plotTotalvsTexturas} left), and contrast ($ r (401) = 0.125 $, $ p = 0.012 $).
Also, a significant difference was found between using textures and not using textures (\textit { $ \chi ^ 2 = 65.017 $, $ p <0.01 $}) (figure \ref{fig:plotTotalvsTexturas} right).

\begin{figure*}  
\centering
\includegraphics[width=0.49\textwidth]{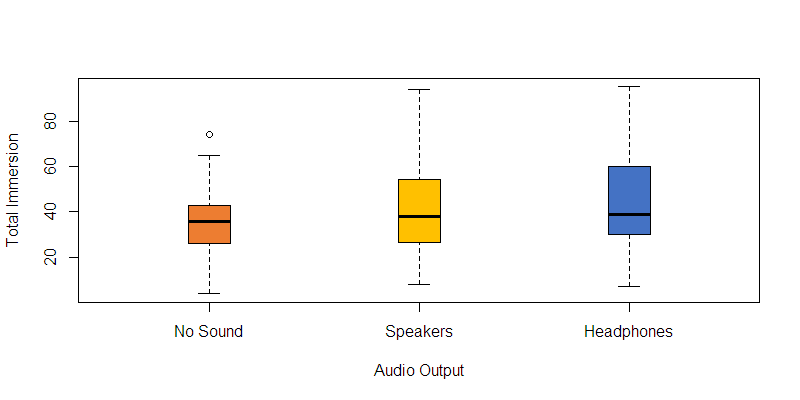}
\includegraphics[width=0.49\textwidth]{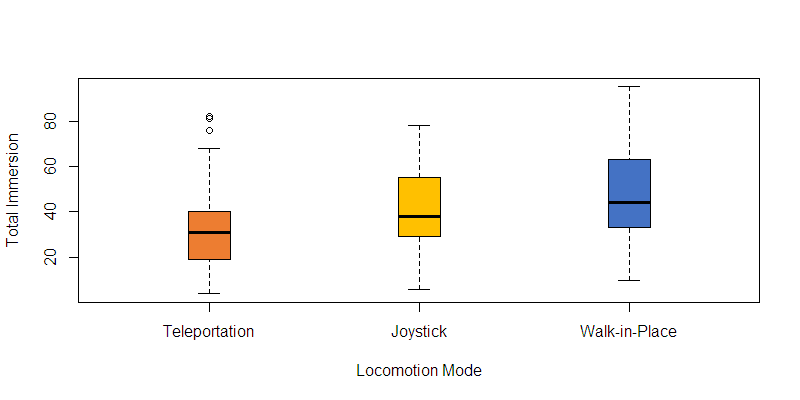}
\caption{
Boxplot for the relationship between audio output and total immersion (\textit{$ \chi ^ 2 = 8.222 $, $ p = 0.02 $}) (left), and between navigation mode and total immersion (\textit {$ \chi ^ 2 = 28.074 $, $ p <0.01 $}) (right).
}
\label{fig:plotTotalvsSalidaAudio}
\end{figure*}

For the audio variables, a statistically significant difference was found between the different audio output modes (\textit {$ \chi ^ 2 = 8.222 $, $ p = 0.02 $}).
According to Dunn's test, this difference is found between the group with no sound and the group with headphones (figure \ref{fig:plotTotalvsSalidaAudio} left).

Regarding the relationship with the locomotion variables, a statistically significant difference was found between the navigation modes (\textit {$ \chi ^ 2 = 28,074 $, $ p <0.01 $}) (figure \ref{fig:plotTotalvsSalidaAudio} right).
Subsequent analysis with Dunn's test revealed that the difference occurs between all groups.

\section{Generation of Models}\label{sec:immersionModel}
In this work, we carried out a process to find the best regression models for immersion based on the 22 independent variables of the experiment.
This process, organized in 3 stages, is described below.

\subsection{\textbf{Stage 1:} Direct Models}\label{sec:stage1}
In the first stage, we fitted regression models using the variables from the experiment.
Five models were generated, detailed in the subsections below and summarized in table \ref{tabla:etapa1}.
Each model is represented by a Total Immersion (TI) function, being $n$ the number of variables ($x$) and $m$ the number of coefficients ($\beta$).

\subsubsection{Simple Linear Model}
This model consists of a linear regression between total immersion and the 22 independent variables.
From these results, we can generate a function of the form:
\begin{equation} 
\footnotesize
\centering
  \begin{aligned}
    TI =& \beta_0 + \beta_1 \times x_1+ \beta_2\times x_2 + \beta_3\times x_3 + ... + \beta_m \times x_n
  \end{aligned}
\end{equation}

\subsubsection{Simple Model with Interactions}
Since the interaction between variables can affect the final result, this multivariate model considers the interactions between each pair of independent variables.
Then, the function corresponding to this model has the following form:

\begin{equation}
\footnotesize
\centering
  \begin{aligned}
    TI =& \beta_0 + \beta_1 \times x_1 + \beta_2 \times x_2 + \beta_3 \times x_1 \times x_2 + ... + \\&
    \beta_m \times x_{n-1} \times x_n
    \end{aligned}
\end{equation}

\subsubsection{Complete Model without Interactions}
In this case, we included all the variables and also the variables of order 2.
The interaction between variables is not considered.
The function corresponding to this model has the following form:

\begin{equation}
\footnotesize
\centering
  \begin{aligned}
    TI =& \beta_0+ \beta_1 \times x_1+ \beta_2 \times x_{1}^2 + \beta_3 \times x_2+ \beta_4 \times x_2^2+ ... + \\&\beta_{m} \times x_n^2
  \end{aligned}
\end{equation}

\subsubsection{Complete Model}
The \textit{Complete Model} is the model that, in addition to including all the variables, includes both the interactions between each pair of independent variables, as well as the order 2 variables.
The function corresponding to this model has the following form:
\begin{equation}
\footnotesize
\centering
  \begin{aligned}
    TI  = & \beta_0 + \beta_1 \times x_1 + \beta_2 \times x_1^2+ \beta_3 \times x_2 + \beta_4 \times x_2^2 + \\
    & \beta_5 \times x_1 \times x_2 + ... + \beta_{m-1} \times  x_{n}^2 + \beta_m \times x_{n-1} \times x_{n}
  \end{aligned}
\end{equation}

\subsubsection{Manual Model}
Generally, researchers rely on theory and experience to decide which candidate variables should be included in a regression model.
In this sense, some techniques recommend that the set of predictor variables included in the final regression model be based on \textit{a priori} data analysis.

In section \ref{sec:dataAnalysis}, we analyzed the statistical relationship between each of the independent variables and the total immersion.
Hence, we propose another model that considers only the variables that affected the total immersion in a statistically significant way.
For this model, we included the variables screen width, frames per second, contrast, duration time, textures, audio output, and navigation mode.
For these variables, we have also included in the model the order 2 variables and the interactions between each pair of independent variables.

\begin{table*}
\centering
\caption{Stage 1 Models Comparison.}

\resizebox{0.7\linewidth}{!}{
\begin{tabular}{lcccc}

\textbf{Model Name} & \multicolumn{1}{l}{$R^2Adjusted$} & \multicolumn{1}{l}{\textbf{AIC}} & \multicolumn{1}{l}{\textbf{Predictors}} & \multicolumn{1}{l}{\textbf{Coefficients}} \\ 
\hline
Simple Linear & 0.4121 & 3303 & 22 & 25 \\ 
\hline
Simple with Interaction & 0.5647 & 3208 & 22 & 299 \\ 
\hline
Complete without Interaction & 0.4423 & 3288 & 22 & 34 \\ 
\hline
Complete & 0.5999 & 3155 & 22 & 308 \\ 
\hline
Manual & 0.4182 & 3289 & 7 & 14 \\
\hline
\end{tabular}}
\label{tabla:etapa1}
\end{table*}

\newcommand{\pvalue}{\mbox{p-value}}

\subsection{\textbf{Stage 2:} Feature Selection}\label{sec:stage2}
Feature selection techniques help to identify a more condensed set of variables that feed the model in a meaningful way.
These techniques iteratively add or remove potential variables, testing for statistical significance after each iteration.

The \textit{Complete Model} presented in the previous section has 308 coefficients and it includes the combination of all the studied variables. 
We performed feature selection to this model as a way to reduce the number of variables, therefore the complexity of the model.
However, decreasing the number of variables of the model can negatively affect the model's predictive power.

The literature presents many feature selection techniques.
We used \textit {Stepwise Regression} because this technique provides the ability to manage large amounts of potential predictor variables, and fine-tuning the model to choose the best predictor variables from the available options.
The Stepwise Regression technique allows us to establish a target \textit{\pvalue}.
The smaller the \textit{\pvalue}, the smaller the number of variables that will fulfill that value, thus obtaining a smaller model.
Hence, we defined 4 groups based on 4 different target \textit{\pvalue{s}}: Model A (\textit{p} = 0.05), Model B (\textit{p} = 0.01), Model C (\textit{p} = 0.005), and Model D (\textit{p} = 0.001).
Finally, for each of these 4 groups, 3 models were generated: one with Forward Selection, another with Backward Selection, and another with Stepwise Selection.
Therefore, 12 new models were obtained, presented in table~\ref{table:etapa2}.

\begin{table*}
\centering
\caption{Stage 2 Models Comparison.}
\resizebox{0.6\linewidth}{!}{%
\begin{tabular}{lcccc} 
\hline
\textbf{Model Name} & \multicolumn{1}{l}{$R^2Adjusted$} & \multicolumn{1}{l}{\textbf{AIC}} & \multicolumn{1}{l}{\textbf{Predictors}} & \multicolumn{1}{l}{\textbf{Coefficients}} \\ 
\hline
A Forward & 0.5999 & 3155 & 22 & 308 \\ 
\hline
A Backward & 0.7604 & 3040 & 22 & 177 \\ 
\hline
A Stepwise & 0.7604 & 3040 & 22 & 177 \\ 
\hline
B Forward & 0.5999 & 3155 & 22 & 308 \\ 
\hline
B Backward & 0.5704 & 3191 & 19 & 40 \\ 
\hline
B Stepwise & 0.5925 & 3172 & 18 & 42 \\ 
\hline
C Forward & 0.5999 & 3155 & 22 & 308 \\ 
\hline
C Backward & 0.492 & 3244 & 13 & 24 \\ 
\hline
C Stepwise & 0.5741 & 3187 & 18 & 39 \\ 
\hline
D Forward & 0.5999 & 3155 & 22 & 308 \\ 
\hline
D Backward & 0.4362 & 3277 & 9 & 15 \\ 
\hline
D Stepwise & 0.4362 & 3277 & 9 & 15 \\
\hline
\end{tabular}
}\label{table:etapa2}
\end{table*}

All the models that used the Forward Selection technique resulted to be equal to the Complete Model from stage 1.
This indicates that the algorithm did not stop until all variables were included.
On the other hand, both the A Backward model and the A Stepwise model, as well as the D Backward model and the D Stepwise model, are also equal to each other.
Two models are equal when they have the same coefficients, predictors and prediction values.

\subsection{\textbf{Stage 3:} Validation}\label{sec:stage3}
We validated all the models created (i.e. the ones from stage 1 and stage 2) using cross-validation with k iterations with repetition.
Ten repetitions were used.
In summary, the k iteration cross-validation procedure with $k = 10$ divides the dataset into 10 subsets.
It uses 9 of these 10 subsets to train the model and the remainder one to test it.
Thus, a prediction error is obtained.
This process is repeated for all the 10 subsets, and the total prediction error is the average of the 10 individual errors.

In this case, we also use repetition, that is, the entire process described above is carried out 10 times.
Hence, the final prediction error is the result of averaging the 10 runs.
This is done for each model, thus obtaining the values of RMSE, $ R ^ 2 $ adjusted, and MAE.

All the models are arranged in table \ref{table:etapa3}, ordered according to the number of coefficients.
This table groups the models that are equal.
As mentioned before, the best prediction can be defined by the highest adjusted $R ^ 2 $ or the lowest RMSE or MAE values.
In this work, we follow the value of $ R ^ 2 $ to decide which model is ``better'' in terms of predictive power.

\begin{table*}
\centering
\caption{Comparison and Validation of all models. The grouped models are exactly the same. The \textbf{Coefficients} column is highlighted to emphasize that the coefficients are sorted from highest to lowest.}
\resizebox{.8\linewidth}{!}{%
\begin{tabular}{ccccc|ccc} 
\hline
\multicolumn{5}{c|}{\textbf{Model Information}} & \multicolumn{3}{c}{\textbf{Validation Information}} \\ 
\hline
\multicolumn{1}{c}{\textbf{Model Name}} &  \multicolumn{1}{l}{\textbf{$R^2 Adjusted$}} & \multicolumn{1}{l}{\textbf{AIC}} & \multicolumn{1}{l}{\textbf{Predictors}} & \multicolumn{1}{l|}{\textbf{Coefficients}}& \multicolumn{1}{l}{\textbf{RMSE}} & \multicolumn{1}{l}{\textbf{$R^2 Adjusted$}} & \multicolumn{1}{l}{\textbf{MAE}} \\ 
\hline
\begin{tabular}[c]{@{}c@{}}Complete\\A Forward\\B Forward\\C Forward\\D Forward\end{tabular}   & 0.5999 & 3155 & 22 & \textbf{308} & 32.18 & 0.1403 & 25.47 \\ 
\hline
\begin{tabular}[c]{@{}c@{}}Simple with\\Interactions\end{tabular}  & 0.5647 & 3208  & 22 & \textbf{299} & 30.56 & 0.1393 & 24.23 \\ 
\hline
\begin{tabular}[c]{@{}c@{}}A Backward\\A Stepwise\end{tabular}   & 0.7604 & 3040 & 22 & \textbf{177} & 12.71 & 0.5973 & 10.23 \\ 
\hline
B Stepwise  & 0.5925 & 3172 & 18 & \textbf{42} & 12.62 & 0.5542 & 10.29 \\ 
\hline
B Backward & 0.5704 & 3191 & 19 & \textbf{40} & 12.99 & 0.5297 & 10.59 \\ 
\hline
C Stepwise  & 0.5741 & 3187 & 18 & \textbf{39} & 12.94 & 0.5314 & 10.68 \\ 
\hline
\begin{tabular}[c]{@{}c@{}}Complete without\\Interactions\end{tabular}  & 0.4423 & 3290 & 22 & \textbf{34} & 14.71 & 0.4037 & 12.01 \\ 
\hline
Simple Linear  & 0.412 & 3303 & 22 & \textbf{25} & 14.86 & 0.3869 & 12.16 \\ 
\hline
C Backward  & 0.492 & 3244 & 13 & \textbf{24} & 13.78 & 0.4691 & 11.36 \\ 
\hline
\begin{tabular}[c]{@{}c@{}}D Backward\\D Stepwise\end{tabular} & 0.4362 & 3277 & 9 & \textbf{15} & 14.36 & 0.4235 & 11.75 \\ 
\hline
Manual & 0.4182 & 3289 & 7 & \textbf{14} & 14.55 & 0.4091 & 11.94 \\
\hline
\end{tabular}
}\label{table:etapa3}
\end{table*}

\begin{table*}
\centering
\caption{Selected Immersion Models.}
\resizebox{0.6\linewidth}{!}{%
\begin{tabular}{ccc|c} 
\hline
\multicolumn{3}{c|}{\textbf{Model Information}} & \textbf{Validation Information} \\ 
\hline
\multicolumn{1}{l}{\textbf{Model Name}} & \multicolumn{1}{l}{\textbf{Coefficients}} & \multicolumn{1}{l|}{\textbf{Predictors}} & $R^2 Adjusted$ \\ 
\hline
Model 1 & 42 & 18 & 0.5542 \\ 
\hline
Model 2 & 24 & 13 & 0.4691 \\ 
\hline
Model 3 & 15 & 9 & 0.4235 \\
\hline
\end{tabular}
}\label{table:modelosFinales}
\end{table*}

\section{Immersion Metrics: Selected Models and Functions}\label{sec:modelosSeleccionados}
From among the obtained models, our goal was to find the one (or ones) that were most closely related to the intended use of the model.
A model with a high predictive power would provide a better immersion approximation based on the variables of the VR system.
A model with fewer predictors requires fewer variables of the VR system.
A model with fewer coefficients can be computed faster.
Hence, when selecting the best models, we need to consider the trade-off between predictive power, number of coefficients, and number of predictors.

Of the resulting models presented in table \ref{table:etapa3}, the model with the best predictive power, based on $ R ^ 2 $, is the A Backward or A Stepwise model, both with 177 coefficients and $ R ^ 2 = 0.5973 $.
The table also presents models with a similar $ R ^ 2 $ and with fewer coefficients.
Therefore, in the search of the best models, we discarded the model with 177 coefficient and analyzed in detail the models with 42, 40, and 39 coefficients that have, respectively, an $ R ^ 2 $ equal to 0.5542, 0.5297, and 0.5314.

Regarding the B Stepwise, B Backward and C Stepwise models, none of them include the variables \textit{reflections}, \textit{reverberation} and \textit{3D spatial sound}.
The B Backward model also does not include the variable \textit{saturation}.
It is interesting that the \textit{3D spatial sound}, which according to the literature is a variable widely influential, was not considered by these models.
On the other hand, some models did not consider the variables \textit{reflections}, \textit{reverberation} and \textit{saturation}.
Taking this into account, we consider that the B Stepwise model, with 42 coefficients, is the best of these three models since, although it has more coefficients, it has greater predictive power.

Then, the Complete Model without Interaction, with 34 coefficients, and the Simple Linear Model, with 25 coefficients, were discarded.
Both include all the 22 variables, and their predictive power is lower than the other models.

We consider that the C Backward Model, with 24 coefficients, is also one of the best models since it includes 13 of the 22 variables, even though its predictive power is lower than the other models with more coefficients and variables.
Of the 22 variables, this model does not include the \textit{stereopsis}, \textit{antialiasing}, \textit{illumination mode}, \textit{saturation}, \textit{shadow strength}, \textit{reflections}, \textit{depth of field}, \textit{reverberation} or \textit{3D spatial sound}.
We consider that the variables \textit{stereopsis}, \textit{lighting mode}, and \textit{3D spatial sound} are relevant since, according to the literature, they have a significant influence on immersion and presence.

\noindent
\begin{equation}
\resizebox{1\linewidth}{!}{$
\begin{aligned}
& \textbf{TotalImmersion}= -52.795864 + \\[-5pt]
&\textit{screenWidth} \times 0.023127 + \\[-5pt]
&\textit{fieldOfView} \times 0.233013 + \\[-5pt]
&\textit{framesPerSecond} \times 1.524708 + \\[-5pt]
&\textit{stereopsisActivated} \times -3.741471 + \\[-5pt]
&\textit{antialiasingActivated} \times -14.463670 + \\[-5pt]
&\textit{textureModeWithTextures} \times -3.558318 + \\[-5pt]
&\textit{illuminationModeLightsAndShading} \times 1.792554 + \\[-5pt]
&\textit{brightness} \times -5.050710 + \\[-5pt]
&\textit{contrast} \times 15.784897 + \\[-5pt]
&\textit{sharpness} \times 26.061258 + \\[-5pt]
&\textit{shadowStrength} \times 14.743503 + \\[-5pt]
&\textit{modelsDetailHigh} \times 1.604808 + \\[-5pt]
&\textit{depthOfFieldActivated} \times 7.430725 + \\[-5pt]
&\textit{particlesActivated} \times 4.942608 + \\[-5pt]
&\textit{audioOutputModeSpeakers} \times 1.870371 + \\[-5pt]
&\textit{audioOutputModeHeadphones} \times 11.826967 + \\[-5pt]
&\textit{ambientSoundActivated} \times 11.696136 + \\[-5pt]
&\textit{locomotionModeJoystick} \times 18.260037 + \\[-5pt]
&\textit{locomotionModeWalkInPlace} \times 2.375830 + \\[-5pt]
&\textit{durationTime} \times 0.002978 + \\[-5pt]
&\textit{screenWidth}^2 \times -0.000008 + \\[-5pt]
&\textit{framesPerSecond}^2 \times -0.017609 + \\[-5pt]
&\textit{screenWidth} \times \textit{illuminationModeLightsAndShading} \times 0.007121 + \\[-5pt]
&\textit{fieldOfView} \times contrast \times -0.227759 + \\[-5pt]
&\textit{fieldOfView} \times sharpness \times -0.431735 + \\[-5pt]
&\textit{stereopsisActivated} \times \textit{shadowStrength} \times -13.691010 + \\[-5pt]
&\textit{stereopsisActivated} \times \textit{durationTime} \times 0.015409 + \\[-5pt]
&\textit{antialiasingActivated} \times \textit{durationTime} \times 0.017004 + \\[-5pt]
&\textit{textureModeWithTextures} \times \textit{contrast} \times 8.744551 + \\[-5pt]
&\textit{textureModeWithTextures} \times \textit{modelsDetailHigh} \times 10.788037 + \\[-5pt]
&\textit{textureModeWithTextures} \times \textit{durationTime} \times 0.019434 + \\[-5pt]
&\textit{shadowStrength} \times \textit{particlesActivated} \times -14.702403 + \\[-5pt]
&\textit{ambientSoundActivated} \times \textit{durationTime} \times -0.017771 + \\[-5pt]
&\textit{locomotionModeJoystick} \times \textit{durationTime} \times -0.014104 + \\[-5pt]
&\textit{locomotionModeWalkInPlace} \times \textit{durationTime} \times 0.016911 + \\[-5pt]
&\textit{modelsDetailHigh} \times \textit{depthOfFieldActivated} \times -8.055026 + \\[-5pt]
&\textit{particlesActivated} \times \textit{ambientSoundActivated} \times 6.549401 + \\[-5pt]
&\textit{antialiasingActivated} \times \textit{audioOutputModeSpeakers} \times 7.244033 + \\[-5pt]
&\textit{antialiasingActivated} \times \textit{audioOutputModeHeadphones} \times -2.993224 + \\[-5pt]
&\textit{antialiasingActivated} \times \textit{illuminationModeLightsAndShading} \times \\[-5pt] &-6.452346 + \\[-5pt]
&\textit{textureModeWithTextures} \times \textit{illuminationModeLightsAndShading} \times \\[-5pt] &-7.164925 \\[-5pt]
\end{aligned}$
\label{eq-modelo1}}
\end{equation}

Finally, the models with 15 and 14 coefficient are very similar in terms of predictive power, number of coefficients, and number of predictors.
The model with 15 coefficients includes 3 variables that the model with 14 coefficients does not.
These are \textit{field-of-view}, \textit{definition} and \textit{models detail}.
On the other hand, the model with 14 coefficients includes a variable that the model with 15 coefficients does not, which is \textit{contrast}.
According to the statistical analysis, the variable \textit{contrast} influences the total immersion, although very slightly.
However, according to the literature, the variables \textit{field-of-view} and \textit{detail of the models} are more significant and influential than \textit{contrast}.
For this reason, we selected the model with 15 coefficients instead of the one with 14.

After this process, three models were selected, which are presented in table \ref{table:modelosFinales}.
For clarity, we will call these models ``Model 1'', ``Model 2'', and ``Model 3''.
The table details the number of coefficients, the number of predictors and the adjusted $ R ^ 2 $, indicating the predictive power of each model.

The functions for Model 1, Model 2 and Model 3 are presented in equations \ref{eq-modelo1}, \ref{eq-modelo2} and \ref{eq-modelo3}, respectively.
These functions can be used to estimate the level of immersion of a given VR system based on its hardware and software features. 

\noindent
\begin{equation}
\resizebox{1\linewidth}{!}{$
\begin{aligned}
& \textbf{Total Immersion}= -38.16095974 +\\[-5pt]
&screenWidth \times 0.008504384 + \\[-5pt]
&fieldOfView \times 0.196812152 + \\[-5pt]
&framesPerSecond \times 1.541130003 + \\[-5pt]
&textureModeWithTextures \times -5.46407892 + \\[-5pt]
&brightness \times -4.085710982 + \\[-5pt]
&contrast \times 19.38644806 + \\[-5pt]
&sharpness \times 23.33455116 + \\[-5pt]
&modelsDetailHigh \times -1.670228672 + \\[-5pt]
&particlesActivated \times -3.069366777 + \\[-5pt]
&audioOutputModeSpeakers \times 5.297975701 + \\[-5pt]
&audioOutputModeHeadphones \times 10.45405873 + \\[-5pt]
&ambientSoundActivated \times -2.809009675 + \\[-5pt]
&locomotionModeJoystick \times 18.95378116 + \\[-5pt]
&ocomotionModeWalkInPlace \times 1.551341478 + \\[-5pt]
&durationTime \times 0.012916461 + \\[-5pt]
&framesPerSecond^2 \times -0.018563525 + \\[-5pt]
&fieldOfView \times contrast \times -0.222598953 + \\[-5pt]
&fieldOfView \times sharpness \times -0.38192204 + \\[-5pt]
&textureModeWithTextures \times modelsDetailHigh \times 8.765077901 + \\[-5pt]
&textureModeWithTextures \times durationTime \times 0.019749586 + \\[-5pt]
&particlesActivated \times ambientSoundActivated \times 7.579476366 + \\[-5pt]
&locomotionModeJoystick \times durationTime \times -0.015178069 + \\[-5pt]
&locomotionModeWalkInPlace \times durationTime \times 0.015704389 \\[-5pt]
\end{aligned}
$\label{eq-modelo2}
}
\end{equation}

\noindent

\begin{equation}
\resizebox{1\linewidth}{!}{%
$
\begin{aligned}
& \textbf{Total Immersion}= -44.78322466 + \\[-5pt]
&screenWidth \times 0.008237546 + \\[-5pt]
&fieldOfView \times 0.227429898 + \\[-5pt]
&framesPerSecond \times 1.608568062 + \\[-5pt]
&textureModeWithTextures \times 9.717910348 + \\[-5pt]
&sharpness \times 26.4102586 + \\[-5pt]
&modelsDetailHigh \times -3.064720396 + \\[-5pt]
&audioOutputModeSpeakers \times 5.519798682 + \\[-5pt]
&audioOutputModeHeadphones \times 10.19128742 + \\[-5pt]
&locomotionModeJoystick \times 5.692683516 + \\[-5pt]
&locomotionModeWalkInPlace \times 13.37931564 + \\[-5pt]
&durationTime \times 0.017860572 + \\[-5pt]
&framesPerSecond^2 \times -0.019223957 + \\[-5pt]
&fieldOfView \times sharpness \times -0.431998513 + \\[-5pt]
&textureModeWithTextures \times modelsDetailHigh \times 9.859037709\\[-5pt]
\end{aligned}$}
\label{eq-modelo3}
\end{equation}

\section{Immersion in Commercial Devices}\label{sec:dispositivosComerciales}
We tested the immersion metrics on three of today's most popular commercial VR systems, with very different hardware and software characteristics each.
These are the Oculus Rift S\footnote{https://www.oculus.com/rift-s/}, the Oculus Quest 2\footnote{https://www.oculus.com/quest/} and the Oculus GO\footnote{https://www.oculus.com/go/}.

To carry out this analysis, we used the application \textit{Beat Saber} which can run on all three devices.
\textit{Beat Saber}\footnote{\label{footnote-label} https://beatsaber.com/} is a rhythm game developed exclusively for VR that has become one of the most popular games in recent years.
The game is developed for the three VR devices we are considering, hence we can use it for the immersion calculation using our metrics.
All three viewers can run the game at 60 frames per second, as indicated by the game specification.

To use our immersion metrics, we need to know the game's software specifications (for example if shadows are being used or not, or the level of brightness or saturation).
Since some of these variables are not specified, we had to estimate them by analyzing gameplays, images, and videos of the game.

The results of the immersion calculation are shown in table \ref{tabla:resultadosPruebasVisores} for the three models and the three devices.
The Oculus GO presented the lowest immersion for the three models.
This is consistent with the technical specifications of the device, as well as users' ratings over the past few years.
This viewer was Oculus' first attempt to make a viewer completely independent of a PC, and its tracking system and visual quality are more basic than the other devices.

On the other hand, the Oculus Rift S and Oculus Quest turned out to be very similar in terms of immersion.
Based on Model 1, the Oculus Rift S presented more immersion than the Oculus Quest, but for Models 2 and 3, the Oculus Quest outperformed the Oculus Rift S.
This small difference between these two devices was expected since, based on the variables used by the metrics, the only difference between the two was the \textit{screen resolution}, which the Oculus Quest narrowly exceeds. 

\begin{table}
\centering
\caption{
Immersion calculated for the three types of devices for the Beat Saber game, using the 3 immersion metrics.}
\resizebox{1\linewidth}{!}{%
\begin{tabular}{cccc} 
\hline
 & \textbf{Oculus Rift S} & \textbf{Oculus Quest} & \textbf{Oculus GO} \\ 
\hline
\textbf{Model 1} & 57.95020 & 54.40024 & 48.17092 \\
\textbf{Model 2} & 60.74711 & 63.46851 & 52.51633 \\
\textbf{Model 3} & 74.07102 & 76.70703 & 58.41741 \\
\hline
\end{tabular}
}
\label{tabla:resultadosPruebasVisores}
\vspace{-0.5 cm}
\end{table}

\section{Discussion}\label{sec:discussion}
The present work studied the different variables of the VR system and how they relate to the perceived level of immersion.
We performed a user study and statistical analyses following a methodology designed to generate immersion metrics.
This section presents a discussion about the obtained results, the limitations of the study, and some directions for future work.

\subsection{Variables}
The statistical analyses described in section \ref{sec:dataAnalysis} provided interesting results.
Some visual variables presented small correlations with immersion, namely the \textit{screen width}, the \textit{frames per second}, and the \textit{contrast}. 
The \textit{screen resolution} and the \textit{frames per second} are variables widely studied in the literature, and it is suggested that a bigger screen resolution and faster frames per second are clearly related to a higher level of immersion.
On the other hand, it was interesting to see that the contrast affected, albeit slightly, the level of immersion.  
This can be related to the role of contrast in detecting the edge and details of objects

The use of textures significantly affected the level of immersion. 
This suggests that the user felt more immersed when the objects and the environment presented a convincing material, similar to what happens in the real world. 
Most objects in the real world present some kind of defined texture or material.
This might explain why a lower immersion was perceived when seeing objects with only solid colors and no textures.

Some visual variables that, according to the literature, significantly affect the immersion, were not relevant in the statistical analysis.
For instance, we expected the \textit{field-of-view} to highly influence the perceived immersion.
Nowadays, every modern VR headset seeks to improve the \textit{field-of-view}, among other variables.
In addition, the \textit{stereopsis} was another variable that did not affect the immersion significantly.
This is a variable directly related to depth perception, both in the real and virtual environment.
It should be considered that there are people who have a deficiency in stereoscopic vision and yet perceive depth. 
This is due to different depth cues in a scene. 
The result obtained is consistent with this and undoubtedly arises when analyzing the different parameters as a whole. 
Therefore, in future work it would be extremely interesting to study the influence and relationship between the variables that provide depth information in more detail.

Regarding the audio, the results are consistent with the literature.
As expected, the use of headphones presented the higher level of immersion, followed by the use of speakers, and the absence of sound.
The headphones deliver the audio to each one of the user's ears, occluding the external noise, and thus improving the immersion, no matter whether the 3D spatial sound, ambient sound, or reverberation were active or not.
It is interesting to note that these three variables did not significantly affected the immersion, since based on the literature, they are relevant. 
The 3D spatial sound, for instance, is not clearly perceived unless the user is wearing headphones.
Future work will consider the analysis of the audio variables in more detail.

Regarding the locomotion mode, there was a clear difference between all groups, being the \textit{walk in place} the most immersive technique, followed by the use of joystick, and finally by teleportation.
In this study, due to physical constrains, the \textit{real walking} technique could not be used. 
However, the results are consistent with the literature, suggesting that the physical body movement of walking did influence the final perceived immersion.

We have relied on the literature and on our previous knowledge to select and study the variables that were used in this work.
However, the study of immersion should not be limited only to these variables.
As technology advances, new variables will emerge that must be considered, studied, and incorporated into the metrics.

\subsection{User Study}
As mentioned in section \ref{sec:participants}, this study was intended to be performed by many participants.
However, due to the conditions of social isolation related to COVID-19, we decided to conduct the experiment focused of a single user. 
Hence, the experiment results would only relate to a target population with the characteristics of the single user who performed the experiment. 
Despite this, the results are still very interesting and, above all, it is extremely important to have a way to consider both the relevance of each variable and the relationships between them in relation to the immersion of the system.
Future work will therefore consider this study by conducting further experiments with more participants.

In this study, we have used a single-item measure to assess immersion. 
There are other questionnaires that provide more information about the different factors that shape presence and immersion but, because they are much larger or complex, participants can get bored and lead to wrong results.
Future work should consider the use other measures to gather more information about the relationship between immersion and the variables of the VR system.

\subsection{Generation of the Metrics}
In this work, we followed a specific methodology to generate immersion metrics, i.e., through the use of regression models in addition to feature selection and validation techniques.
Other alternatives or techniques can be considered in the different parts of the metrics generation process to get more insight about the relationship between the variables and the effect on the immersion.
We have made the dataset available online for the public.
Future work, therefore, should consider the study and application of other techniques.

After generating the immersion models, a selection process was carried out to determine which one (or ones) of these could be considered the best models.
In this process (described in section \ref{sec:modelosSeleccionados}), we made decisions to discard some models in favor of others. 
For this purpose, we focused on the predictive power, the number of coefficients and the number of predictors of the models, without considering the particular variables of each model.
However, for a particular system or application, it might be interesting to favor the model that includes a particular variable such as, for example, \textit{3D spatial audio}.
This should be considered in future work

Based on these results, the Complete Model turned out to be not as powerful as it seemed on stage 1, now obtaining only a $ R ^ 2 = 0.1403 $. 
This could most likely be due to \textit{overfitting}.
It is highly probable that a model that uses all the variables and all the combinations between them will be adjusted to very specific characteristics of the training data that have no causal relationship with the objective function.

Our immersion metrics are intended to work with any VR system, based on its hardware and software characteristics.
However, as mentioned above, some of the studied variables depend on both the virtual scenario being used, as well as the specific task being performed. 
In this sense, future work should also consider the evaluation of immersion metrics in various case studies and different application domains.

\section{Conclusions}\label{sec:conclusions}
Currently, the development of new VR systems with different hardware and software characteristics has been accelerated.
Every system tries to outperform the others, but most of them rely only on technological advances to improve the user's immersion and experience.
However, not only the most common hardware variables (such as the field-of-view or the screen resolution) should be considered.

VR systems consider both hardware and software variables that influence the total immersion of the system.
It is necessary to know which variables are most influential and how.
Thus, for example, if we need to select among different variables to include in a new VR system, we may choose those with the highest impact on the level of immersion.
The influence of these hardware and software variables on immersion has only been considered individually or taking into account small groups of these.
To date, the way they all simultaneously affect immersion has not been analyzed. 
The motivation of this study is based on the study and application of these hardware and software variables of the VR system and their relationship to construct an immersion metric.
In this way, the level of immersion of any VR system can be estimated without the need of user tests.

The work we carried out has been highly challenging.
The obtained results contribute to the area of immersive technologies and more specifically to the area of VR.
Commercial VR systems developed in recent years are based on the assumption that the better the hardware the higher the immersion and therefore, the better the experience.
Even though upgrading the hardware can help to improve immersion, this is not the only issue to be considered. 
To truly improve immersion, the combination of variables to be considered must be improved.
Immersion metrics can be designed to consider these characteristics of a VR system and help to decide which variables to favor both when designing a new VR system and when estimating the immersion of existing VR systems.
This allows the comparison between different systems, being able to choose the best alternative according to the task to be performed.

\bibliographystyle{unsrt}
\bibliography{references}

\end{document}